\documentclass[aps,preprint]{revtex4}
\usepackage[dvips]{graphicx}
\begin{document}
\draft
\title{\bf Experimental and numerical investigation of the reflection coefficient and the distributions of Wigner's
reaction matrix for irregular graphs with absorption}

\author{Micha{\l} {\L}awniczak$^1$, Oleh Hul$^1$, Szymon Bauch$^1$, Petr \v{S}eba$^{2,3}$, and Leszek Sirko$^1$}
\address{$^1$Institute of Physics, Polish Academy of Sciences,
Aleja \ Lotnik\'{o}w 32/46, 02-668 Warszawa, Poland \\
$^2$University of Hradec Kr\'alov\'e, Hradec
Kr\'alov\'e, Czech Republic\\
$^3$Institute of Physics, Academy of Sciences of the Czech Republic,
Cukrovarnick\'{a} 10, 162 53 Praha, Czech Republic}

\date{April 16, 2008}

\bigskip

\begin{abstract}
We present the results of experimental  and numerical study of the
distribution of the reflection coefficient $P(R)$ and the
distributions of the imaginary $P(v)$ and the real $P(u)$ parts of
the Wigner's reaction $K$ matrix for irregular fully connected
hexagon networks (graphs) in the presence of strong absorption. In
the experiment we used microwave networks, which were built of
coaxial cables and attenuators connected by joints. In the
numerical calculations experimental networks were described by
quantum fully connected hexagon graphs. The presence of absorption
introduced by attenuators was modelled by optical potentials.  The
distribution of the reflection coefficient $P(R)$ and the
distributions of the reaction $K$ matrix were obtained from the
measurements and numerical calculations of the scattering matrix
$S$ of the networks and graphs, respectively. We show that the
experimental and numerical results are in good agreement with the
exact analytic ones obtained within the framework of random matrix
theory (RMT).
\end{abstract}

\bigskip
\maketitle

\smallskip

Quantum graphs of connected one-dimensional wires  were introduced
more than sixty years ago by Pauling \cite{Pauling}. Next the same
idea was used  by Kuhn \cite{Kuhn}  to describe organic molecules
by free electron models. Quantum graphs can be considered as
idealizations of physical networks in the limit where the lengths
of the wires are much bigger than their widths, i.e. assuming that
the propagating waves remain in a single transversal mode. Among
the systems modelled by quantum graphs one can find e.g.,
electromagnetic optical waveguides \cite{Flesia, Mitra},
mesoscopic systems \cite{Imry, Kowal} , quantum wires
\cite{Ivchenko, Sanchez} and excitation of fractons in fractal
structures \cite{Avishai, Nakayama}. Recently it has been shown
that quantum graphs are excellent paradigms of quantum chaos
\cite{Kottossmilansky,Kottos,Prlkottos,Zyczkowski,Kus,Tanner,
Kottosphyse,Kottosphysa,Gaspard,Blumel,Hul2004}. More complicated
and thus more realistic systems - microwave networks with moderate
absorption strength $\gamma =2\pi \Gamma /\Delta \leq 7.1$, where
$\Gamma$ is the absorption width and $\Delta$ is the mean level
spacing, have been experimentally investigated in
\cite{Hul2005,Hul2007}. Other interesting open objects - quantum
graphs with leads - have been analyzed in details in
\cite{Kottosphyse,Kottosphysa}. However, the properties of
networks and graphs with strong absorption have not been studied
experimentally neither numerically so far. Therefore, in this
paper we study experimentally and numerically the distribution of
the reflection coefficient $P(R)$ and the distributions of the
Wigner's reaction matrix \cite{Akguc2001} (in the literature often
called $K$ matrix \cite{Fyodorov2004}) for networks (graphs) with
time reversal symmetry ($\beta=1$) in the presence of strong
absorption.

In the case of a single channel antenna experiment the $K$ matrix
is related to the scattering matrix $S$ by the following relation
\begin{equation}
\label{Eq.1} S=\frac{1-iK}{1+iK}.
\end{equation}
Eq. (1) holds for the systems with absorption but without direct
processes \cite{Fyodorov2004}. It is important to mention that the
function $Z=iK$ has a direct physical meaning of the electric
impedance that has been recently measured in the microwave cavity
experiment \cite{Anlage2005}. In the one channel case the $S$
matrix can be parameterized as
\begin{equation}
\label{Eq.2} S=\sqrt{R}e^{i\theta},
\end{equation}
where $R$ is the reflection coefficient and $\theta$ the phase.

Properties of the statistical distributions of the $S$ matrix with
direct processes and imperfect coupling have been studied
theoretically in several important papers
\cite{Lopez1981,Doron1992,Brouwer1995,Savin2001,Fyodorov2003,Fyodorov2005}.
Recently the distribution of the $S$ matrix has  been also
measured experimentally for chaotic microwave cavities with
absorption \cite{Kuhl2005}. The distribution $P(R)$ of the
reflection coefficient $R$ and the distributions of the imaginary
$P(v)$ and the real $P(u)$ parts of the Wigner's reaction $K$
matrix are theoretically known for any dimensionless absorption
strength $\gamma $ \cite{Fyodorov2004,Savin2005}. In the case of
time reversal systems (symmetry index $\beta=1$) $P(R)$ has been
studied experimentally by M\'endez-S\'anchez et al.
\cite{Sanchez2003}. The distributions $P(v)$ and $P(u)$ have been
studied for chaotic microwave cavities in
\cite{Anlage2005,Anlage2006} and for microwave networks for
moderate absorption strength $\gamma \leq 7.1$ in
\cite{Hul2005,Hul2007}. For systems without time reversal symmetry
($\beta=2$) and a single perfectly coupled channel $P(R)$ was
calculated  by Beenakker and Brouwer \cite{Beenakker2001} while
the exact formulas for the distributions $P(v)$ and $P(u)$ were
given by Fyodorov and Savin \cite{Fyodorov2004}.

In the experiment quantum graphs can be simulated by microwave
networks. The analogy between quantum graphs and microwave
networks is based upon the equivalency of the Schr\"odinger
equation describing the quantum system and the telegraph equation
describing the microwave circuit \cite{Hul2004}.

A general microwave network consists of $N$ vertices connected by
bonds e.g., coaxial cables. A coaxial cable consists of an inner
conductor of radius $r_1$ surrounded by a concentric conductor of
inner radius $r_2$. The space between the inner and the outer
conductors is filled with a homogeneous material having a
dielectric constant $\varepsilon$. For a frequency $\nu$ below the
onset of the next TE$_{11}$ mode only the fundamental TEM mode can
propagate inside a coaxial cable. (This mode is in the literature
often called a Lecher wave.)  The cut-off frequency of the
TE$_{11}$ mode is $\nu_{c} \simeq \frac{c}{\pi (r_1+r_2)
\sqrt{\varepsilon}} = 32.9$ GHz \cite{Jones}, where $r_1$ = 0.05
cm is the inner wire radius of the coaxial cable (SMA-RG402),
while $r_2$ = 0.15 cm is the inner radius of the surrounding
conductor, and $\varepsilon \simeq 2.08$ is the Teflon dielectric
constant \cite{Breeden1967,Savytskyy2001}.

From the experimental point of view absorption of the networks can
be changed by the change of the bonds' (cables') lengths
\cite{Hul2004} or more effectively by the application of microwave
attenuators \cite{Hul2005,Hul2007}. In the numerical calculations
weak absorption inside the cables can be described with the help
of complex wave vector \cite{Hul2004}. We will show that strong
absorption inside an attenuator can be  described by a simple
optical potential. The corresponding mathematical theory has been
developed in \cite{Ex1}.

The distribution $P(R)$ of the reflection coefficient $R$ and the
distributions of the imaginary and real parts of the Wigner's
reaction matrix $K$ for microwave networks with absorption were
found using the impedance approach
\cite{Anlage2005,Anlage2006,Hul2007}. In this approach the real and
imaginary parts of the normalized impedance $Z$
\begin{equation}
\label{Equation22} Z = \frac{\textrm{Re } Z_{n}+i(\textrm{Im }
Z_{n}-\textrm{Im } Z_{r})}{\textrm{Re } Z_{r}}
\end{equation}
of a chaotic microwave system are measured, with
$Z_{n(r)}=Z_0(1+S_{n(r)})/(1-S_{n(r)})$ being the network
(radiation) impedance expressed by the network (radiation)
scattering matrix $S_{n(r)}$ and $Z_0$ is the characteristic
impedance of the transmission line. The radiation impedance
$Z_{r}$ is the impedance seen at the output of the coupling
structure for the same coupling geometry, but with the vertices of
the network removed to infinity. The Wigner's reaction matrix $K$
can be expressed by the normalized impedance as $K=-iZ$. The
scattering matrix $S$ of a network  for the perfect coupling case
(no direct processes present) required for the calculation of the
reflection coefficient $R$ (see Eq. (2)) can be finally extracted
from the formula $S=(1-Z)/(1+Z)$.

\begin{figure}[!]
\begin{center}
\rotatebox{270} {\includegraphics[width=0.5\textwidth,
height=0.8\textheight, keepaspectratio]{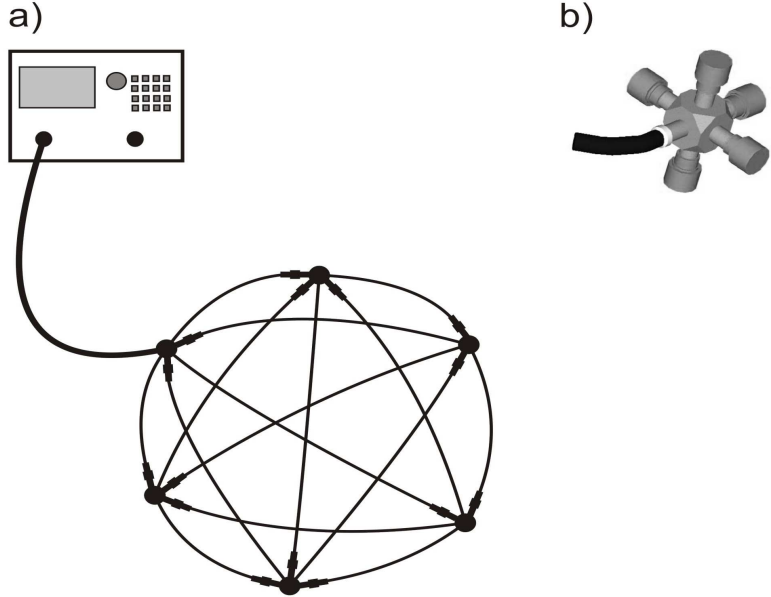}} \caption{(a) The
scheme of the experimental set-up for measurements of the
scattering matrix $S_n$ of the microwave fully connected networks
with absorption. Absorption in the networks was varied by the
change of the attenuators. (b) The scheme of the setup used to
measure the radiation scattering matrix $S_{r}$. Instead of a
network five 50 $\Omega $ loads were connected to the
6-joint.}\label{Fig1}
\end{center}
\end{figure}

Figure~1(a) shows the experimental setup for measuring  the
single-channel scattering matrix $S_{n}$ of fully connected hexagon
microwave networks necessary for finding of the impedance $Z_{n}$.
We used Hewlett-Packard 8720A microwave vector network analyzer to
measure the scattering matrix $S_n$ of the networks in the frequency
window: 7.5--11.5 GHz.  The networks were connected to the vector
network analyzer through a lead - a HP 85131-60012 flexible
microwave cable  - connected to a 6-joint vertex. The other five
vertices of the networks were connected by  5-joints. Each bond of
the network presented in Fig.~1(a) contains a microwave attenuator.

The radiation impedance $Z_{r}$ was found experimentally by
measuring the scattering matrix $S_{r}$ of the 6-joint connector
with five joints terminated by 50 $\Omega $ loads (see
Figure~1(b)).

The experimentally measured fully connected hexagon networks were
described in numerical calculations by quantum  fully connected
hexagon graphs with one lead attached to the 6-joint vertex. In
the calculations  attenuators (absorbers) were modelled by optical
potentials \cite{Ex1}. To be explicit we suppose that the fully
connected hexagon graph $\Upsilon$ with one coupled antenna is
described in the Hilbert space $L^2(\Upsilon):= \bigoplus_{(j,n)}
L^2(0,\ell_{jn})\bigoplus L^2(0,\infty)$, where $\ell_{jn}$ stays
for the lengths of the bond connecting the vertices $j$ and $n$
and the halfline $(0,\infty)$ describes the attached antenna.

We define the Schr\"odinger operator $H$ by
\begin{equation} \label{graph SO}
H{\psi_{jn} := \, -\psi''_{jn}+ U_{jn}\psi_{jn}},
\end{equation}

with $\psi_{jn}\in L^2(0,\ell_{jn})$ for the bonds and

\begin{equation} \label{graph ant}
H{\psi_{0n} := \, -\psi_{0n}''},
\end{equation}

with $\psi_{0n}\in L^2(0,\infty)$ describing the wave function of
the antenna connected to the vertex $n$ (note that the "infinite"
vertex of the antenna has index 0) .

At the vertices the wave functions are linked together with the
boundary values

\begin{equation} \label{boundary values}
\psi_{jn}(j):=\lim_{x\to 0+} \psi_{jn}(x)\,, \quad \psi'_{jn}(j):\lim_{x\to 0+} \psi'_{jn}(x)\,,
\end{equation}

satisfying  boundary conditions
$\,\psi_{jn}(j)=\psi_{jm}(j)=:\psi_j$ for all $n,m$ describing
connected vertices, and

\begin{equation} \label{delta}
\sum_{n\in\nu(j)} \psi'_{jn}(j)= 0.
\end{equation}

The optical potentials $U_{jn}$ are purely imaginary and describe
the absorber inserted between the vertices $(j,n)$.

Since the graph $\Upsilon$ is infinite (due to the attached
antenna) we can look  for solutions of the equation

\begin{equation} \label{local SO}
H\psi= k^2\psi,
\end{equation}

referring to the continuous spectrum, where $k$ is the wave
vector. For microwaves propagating inside  a lossless bond with a
dielectric constant $\varepsilon$ the wave vector $k=2\pi
\varepsilon \nu/c $, where $\nu $ and $c$ denote the frequency of
a microwave field and the speed of light in the vacuum,
respectively. To solve this equation we used the graph duality
principle \cite{Ex1}. According to this principle we need to solve
the equation $-f''+U_{jn}f=k^2f$ on $[0,\ell_{jn}]$ satisfying the
normalized Dirichlet boundary conditions

\begin{equation}
u_{jn}(\ell_{jn})= 1\!-\!(u_{jn})'(\ell_{jn})=0\,, \;\; v_{jn}(0)1\!-\!(v_{jn})'(0)=0\,.
\end{equation}

The Wronskian of this solution is naturally equal to $W_{jn}-v_{jn}(\ell_{jn}) =u_{jn}(0)$. Then according to \cite{Ex1} the
corresponding boundary values (\ref{boundary values}) satisfy the
equation

\begin{equation} \label{discrete delta}
\sum_{n} {\psi_n\over W_{jn}}\,-\, \left(\, \sum_{n\in\nu(j)}
{(v_{jn})'(\ell_{jn}) \over W_{jn}}\, \right)\psi_j\,=\, 0\,.
\end{equation}

Conversely, any solution $\psi_j$ of the system  (\ref{discrete
delta}) determines a solution of (\ref{local SO}) by

\begin{eqnarray}
\psi_{jn}(x)= {\psi_n\over W_{jn}}\,u_{jn}(x) -\,{\psi_j\over
W_{jn}}\,v_{jn}(x) \;\;
& {\rm if} & n = 1,..,6\,, \label{reconstruction i} \\
\psi_{jn}(x)= -\,{\psi_j\over W_{jn}}\,v_{jn}(x) \;\; & {\rm if} &
n=0\,. \label{reconstruction b}
\end{eqnarray}

As already mentioned the microwave attenuators are modelled by
optical potentials localized inside the inserted component. It is
well known that any smooth and localized potential can be easily
approximated by a sequence of delta potentials inside the support
of the potential - see \cite{demkov} for details. We will use this
fact and express the optical potential as a sum of $N$
delta-potentials with imaginary coupling constants:
$U(x)=ib\sum_{r=1}^{N} \delta(x-(r-1)l_b/(N-1))$. The
delta-potentials were equally spaced inside the length $l_{b}$ of
the absorbing element (attenuator). By changing the number $N$ and
the strength $b$ of delta-potentials we were able to vary
absorbing properties as well as reflective properties of
attenuators. We used $N=10$ delta-potentials with $b=0.028$
$m^{-1}$ for simulation of the 1 dB attenuators and $N=12$
delta-potentials with $b=0.045$ $m^{-1}$ for the 2 dB attenuators.
In both cases the length of the attenuator was $l_{b}=2.65$ cm.
Furthermore, in the numerical calculations of the scattering
matrices $S_n$ of the graphs the weak absorption inside the
microwave cables was taken into account by replacing the real wave
vector $k$ by the complex vector $k+ia\sqrt{k}$ \cite{Goubau},
where the absorption coefficient was assumed to be $a=0.009$
$m^{-1/2}$ \cite{Hul2004}.

\begin{figure}[!]
\begin{center}
\rotatebox{270} {\includegraphics[width=0.5\textwidth,
height=0.8\textheight, keepaspectratio]{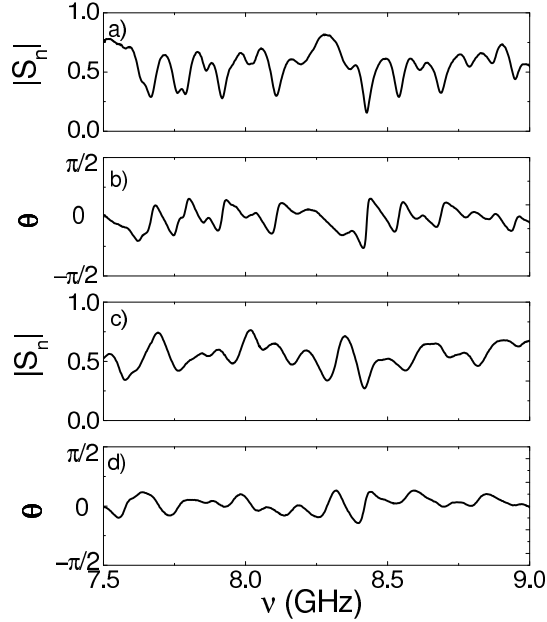}} \caption{In the
panels (a) and (b) the modulus $|S_n|$ and the phase $\theta $ of
the scattering matrix $S$ measured for the network with $\gamma =
19.9$ are plotted in the frequency range 7.5 - 9 GHz. In (c) and
(d) $|S_n|$ and $\theta $ of the scattering matrix $S_n$ are
plotted for the network with $\gamma = 47.9$ in the same frequency
range. The measurements have been done for the two networks which
in each bond contained: 1 dB attenuator ((a) and (b)) and 2 dB
attenuator ((c) and (d)), respectively. The total ``optical"
length of the microwave networks including joints and attenuators
were 574 cm and 554 cm, respectively.}\label{Fig2}
\end{center}
\end{figure}

In order to find the distribution $P(R)$ of the reflection
coefficient $R$ and the distributions of the imaginary and real
parts of the K matrix we measured the scattering matrix $S_n$ of
$88$ and $74$  network configurations containing in each bond a
single 1 dB and 2 dB microwave SMA attenuator, respectively. The
total optical lengths of the microwave networks containing 1 dB
attenuators, including joints and attenuators, varied from 574 cm
to 656 cm. For the networks with 2 dB attenuators the optical
lengths varied from 554 cm to 636 cm. To avoid degeneracy of
eigenvalues of the networks the lengths of the bonds were chosen
as incommensurable.

In Figure~2 the modulus $|S_n|$ and the phase $\theta $ of the
scattering matrix $S_n$ of the microwave networks with $\gamma =
19.9 \textrm{ and } 47.9$, respectively, are presented in the
frequency range 7.5 - 9 GHz. The measurements were done for two
networks containing 1 dB and 2 dB attenuators, respectively. Their
total ``optical" lengths including joints and attenuators were 574
cm and 554 cm, respectively.

For systems with time reversal symmetry ($\beta=1$), the explicit
analytic expression for the distribution $P(R)$ of the reflection
coefficient $R$ is given by \cite{Savin2005}

\begin{equation}
\label{Equation23}
P(R)=\frac{2}{(1-R)^2}P_0\Bigl(\frac{1+R}{1-R}\Bigr).
\end{equation}

The probability distribution $P_0(x)$ is given by the
expression

\begin{equation}
P_0(x)=-\frac{dW(x)}{dx},
\end{equation}

where the integrated probability distribution $W(x)$ is expressed by the
formula \cite{Savin2005}

\begin{equation}
W(x)=\frac{x+1}{4\pi}\Bigl[f_{1}(w)g_{2}(w)+f_{2}(w)g_{1}(w)+h_{1}(w)j_{2}(w)+h_{2}(w)j_{1}(w)\Bigr]_{w=(x-1)/2}.
\end{equation}

The functions $f_{1},g_{1},h_{1},j_{1}$  are defined as follows

\begin{equation}
f_{1}(w)=\int_{w}^{\infty}dt\frac{\sqrt{t\mid t-w\mid}e^{-\gamma
t/2}}{(1+t)^{3/2}}\Bigl[1-e^{-\gamma}+\frac{1}{t}\Bigr],
\end{equation}

\begin{equation}
g_{1}(w)=\int_{w}^{\infty}dt\frac{e^{-\gamma t/2}}{\sqrt{t\mid
t-w\mid}(1+t)^{3/2}},
\end{equation}

\begin{equation}
h_{1}(w)=\int_{w}^{\infty}dt\frac{\sqrt{\mid t-w\mid} e^{-\gamma
t/2}}{\sqrt{t(1+t)}}\Bigl[\gamma+(1-e^{-\gamma})(\gamma
t-2)\Bigr],
\end{equation}

\begin{equation}
j_{1}(w)=\int_{w}^{\infty}dt\frac{e^{-\gamma t/2}}{\sqrt{t\mid
t-w\mid} (1+t)^{1/2}}.
\end{equation}

Their counterparts with the index 2 are given by the same
expressions but the integration is performed in the interval $t \in
[0, w]$ instead of $[w, \infty)$.

\begin{figure}[!]
\begin{center}
\rotatebox{270} {\includegraphics[width=0.5\textwidth,
height=0.6\textheight, keepaspectratio]{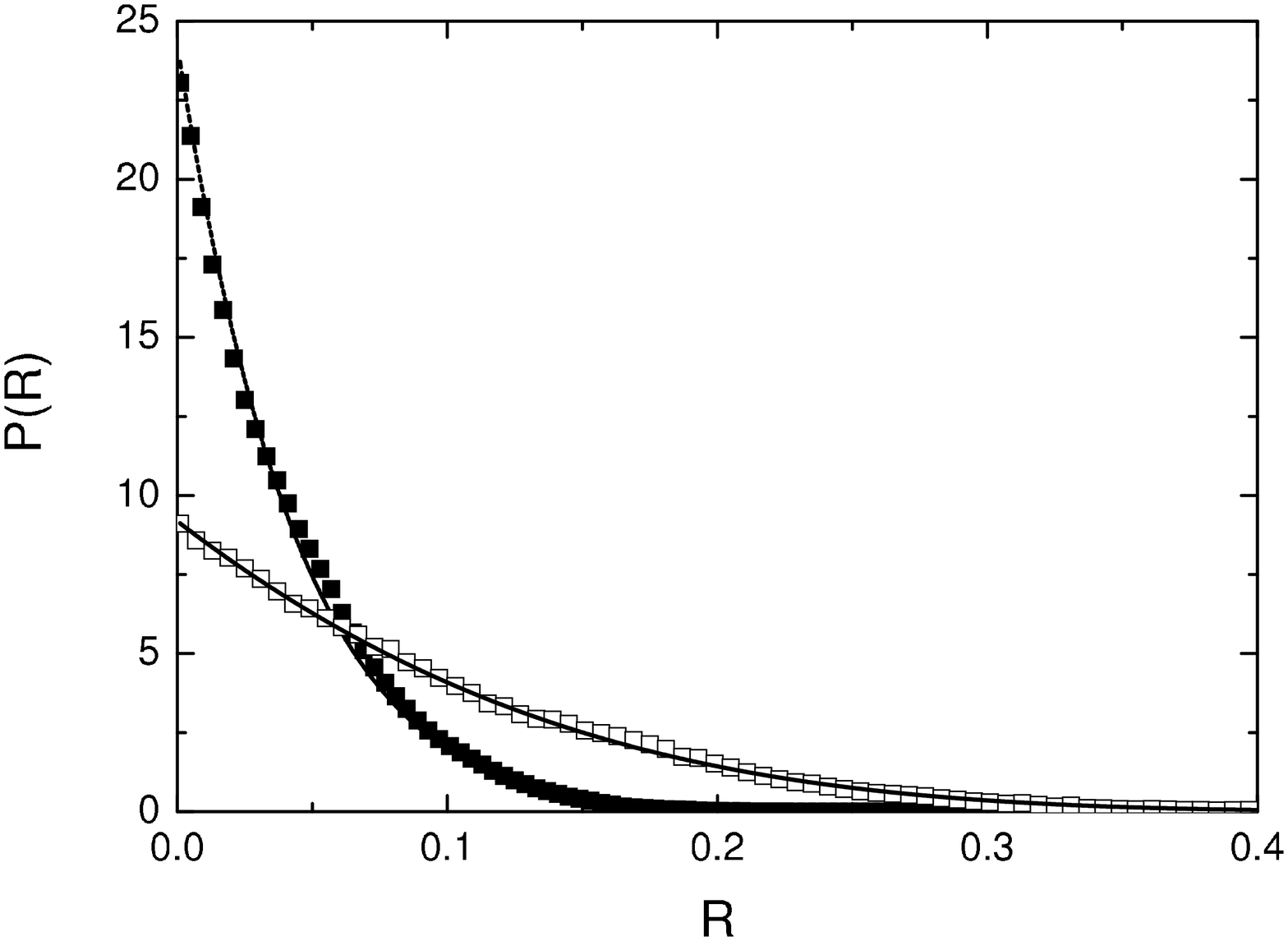}}
\caption{Experimental distribution $P(R)$ of the reflection
coefficient $R$ for the microwave fully connected hexagon networks
at $\bar{\gamma } = 19.3$ (open squares) and $\bar{\gamma } =
47.7$ (full squares). The corresponding theoretical distribution
$P(R)$ evaluated from the Eq.~(\ref{Equation23}) is marked by the
solid line ($\gamma = 19.3$) and dashed line ($\gamma = 47.7$),
respectively.}\label{Fig3}

\rotatebox{270} {\includegraphics[width=0.5\textwidth,
height=0.6\textheight, keepaspectratio]{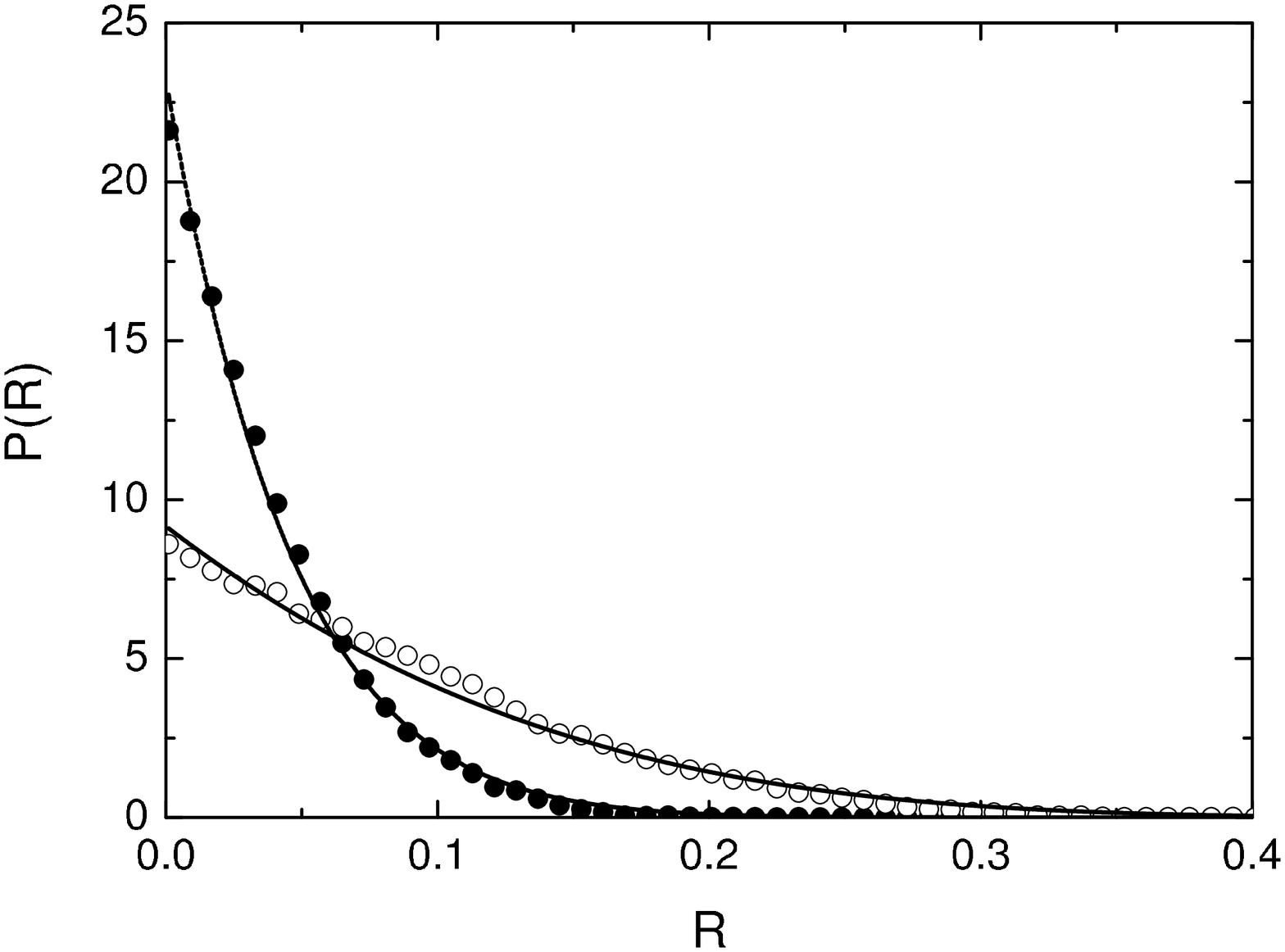}} \caption{Numerical
distribution $P(R)$ of the reflection coefficient $R$ for fully
connected hexagon quantum graphs at $\bar{\gamma } = 19.3$ (open
circles) and $\bar{\gamma } = 47.7$ (full circles).  The
corresponding theoretical distribution $P(R)$ evaluated from the
Eq.~(\ref{Equation23}) is marked by the solid line ($\gamma =
19.3$) and dashed line ($\gamma = 47.7$),
respectively.}\label{Fig4}
\end{center}
\end{figure}

The distributions of the imaginary and  the real parts $P(v)$ and
$P(u)$ of the $K$ matrix \cite{Fyodorov2004} can be also expressed
by the probability distribution $P_0(x)$:
\begin{equation}
\label{Equation34} P(v)=\frac{\sqrt{2}}{\pi
v^{3/2}}\int^{\infty}_{0}dqP_0\Bigl[q^2+\frac{1}{2}\Bigl(v+\frac{1}{v}\Bigr)\Bigr],
\end{equation}
and
\begin{equation}
\label{Equation35} P(u)=\frac{1}{2\pi
\sqrt{u^{2}+1}}\int^{\infty}_{0}dqP_0\Bigl[\frac{\sqrt{u^2+1}}{2}\Bigl(q+\frac{1}{q}\Bigr)\Bigr],
\end{equation}
where $-v=\textrm{Im} \, K<0$ and $u=\textrm{Re} \,K$ are,
respectively, the imaginary and real parts of the $K$ matrix.

Figure~3 shows the experimental distributions $P(R)$ (squares) of
the reflection coefficient $R$ for two mean values of the
parameter $\bar{\gamma }$, viz., 19.3 and 47.7. The distribution
for $\bar{\gamma } = 19.3$ is obtained by averaging over 88
realizations of the microwave networks containing 1 dB
attenuators. The distribution for $\bar{\gamma } = 47.7$ is
obtained by averaging over 74 realizations of the microwave
networks containing 2 dB attenuators. The experimental values of
the $\gamma$ parameter were estimated for each realization of the
network by adjusting the theoretical mean reflection coefficient
$\langle R \rangle _{th}$ to the experimental one $\langle R
\rangle=\langle SS^{\dag}\rangle $, where
\begin{equation}
\label{Equation36} \langle R \rangle _{th} = \int _0^1dRRP(R).
\end{equation}

Figure~3 also presents the  corresponding distributions $P(R)$
(solid and dashed lines, respectively) evaluated from
Eq.~(\ref{Equation23}). A good overall agreement of the
experimental distributions $P(R)$ with their theoretical
counterparts is seen.

Figure~4 shows the numerically evaluated distributions $P(R)$
(circles) of the reflection coefficient $R$ for the graphs at
$\bar{\gamma } = 19.3 \textrm{ and } 47.7$ compared to the
theoretical ones evaluated from the formula
Eq.~(\ref{Equation23}). The numerical distributions are the result
of averaging over 162 and 214 realizations of the graphs with
optical potentials simulating 1 dB and 2 dB attenuators,
respectively. The numerical values of $\gamma$ parameter were also
estimated by adjusting the theoretical mean reflection coefficient
to the numerical one. The agreement between the numerical results
for $\bar{\gamma } = 47.7$ and the theoretical ones (dashed line)
is good. However, for $\bar{\gamma } = 19.3$ for $R < 0.15$ some
discrepancies between the numerical results and the theoretical
ones (solid line) are visible.

\begin{figure}[!]
\begin{center}
\rotatebox{270} {\includegraphics[width=0.5\textwidth,
height=0.6\textheight, keepaspectratio]{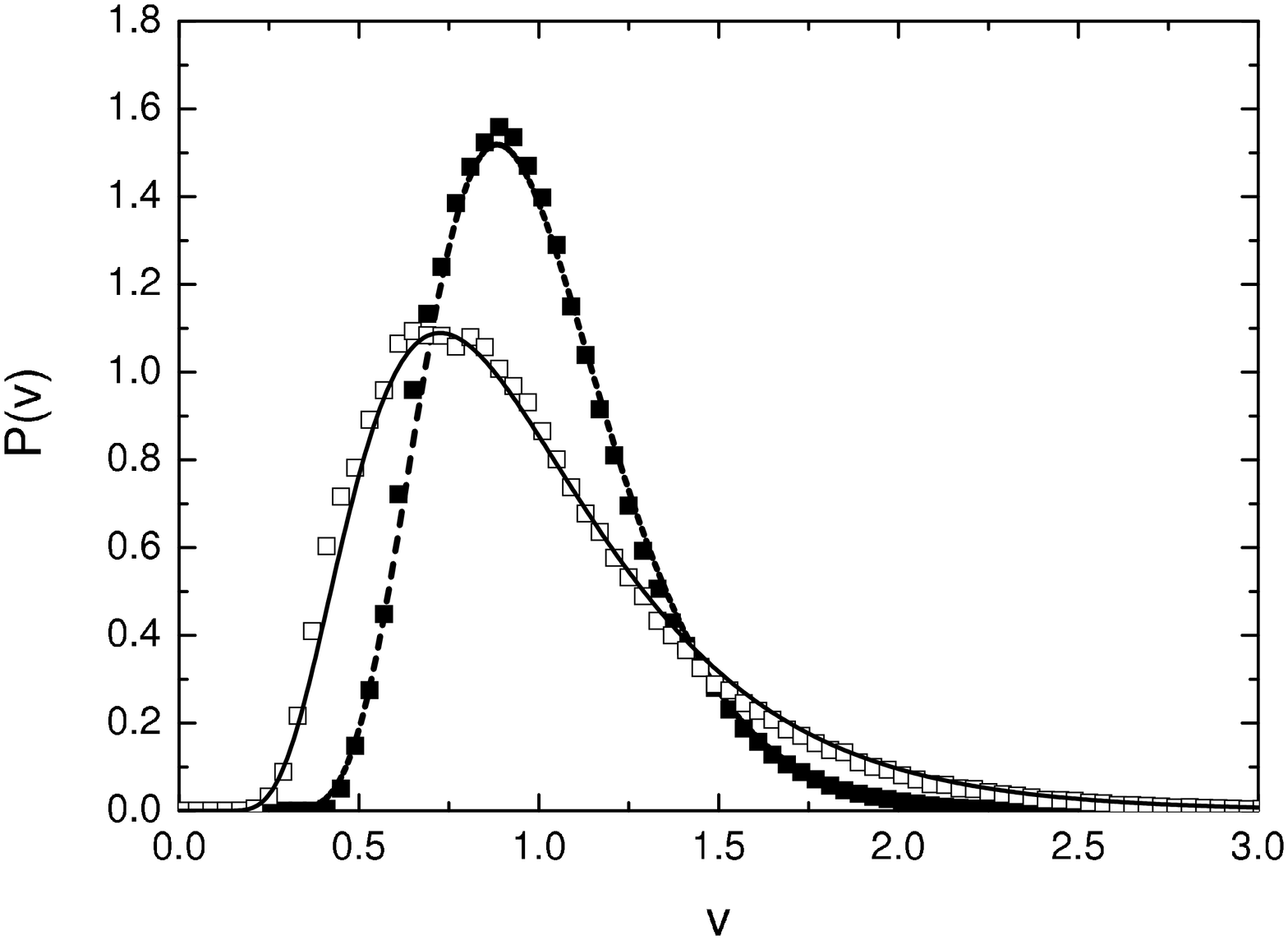}}
\caption{Experimental distribution $P(v)$ of the imaginary part of
the $K$ matrix for the two values of the mean absorption
parameter: $\bar{\gamma } = 19.3$ (open squares) and $\bar{\gamma
} = 47.7$ (full squares), respectively. The corresponding
theoretical distribution $P(v)$ evaluated from the
Eq.~(\ref{Equation34}) is marked by the solid line ($\gamma =
19.3$) and dashed line ($\gamma = 47.7$),
respectively.}\label{Fig5}

\rotatebox{270} {\includegraphics[width=0.5\textwidth,
height=0.6\textheight, keepaspectratio]{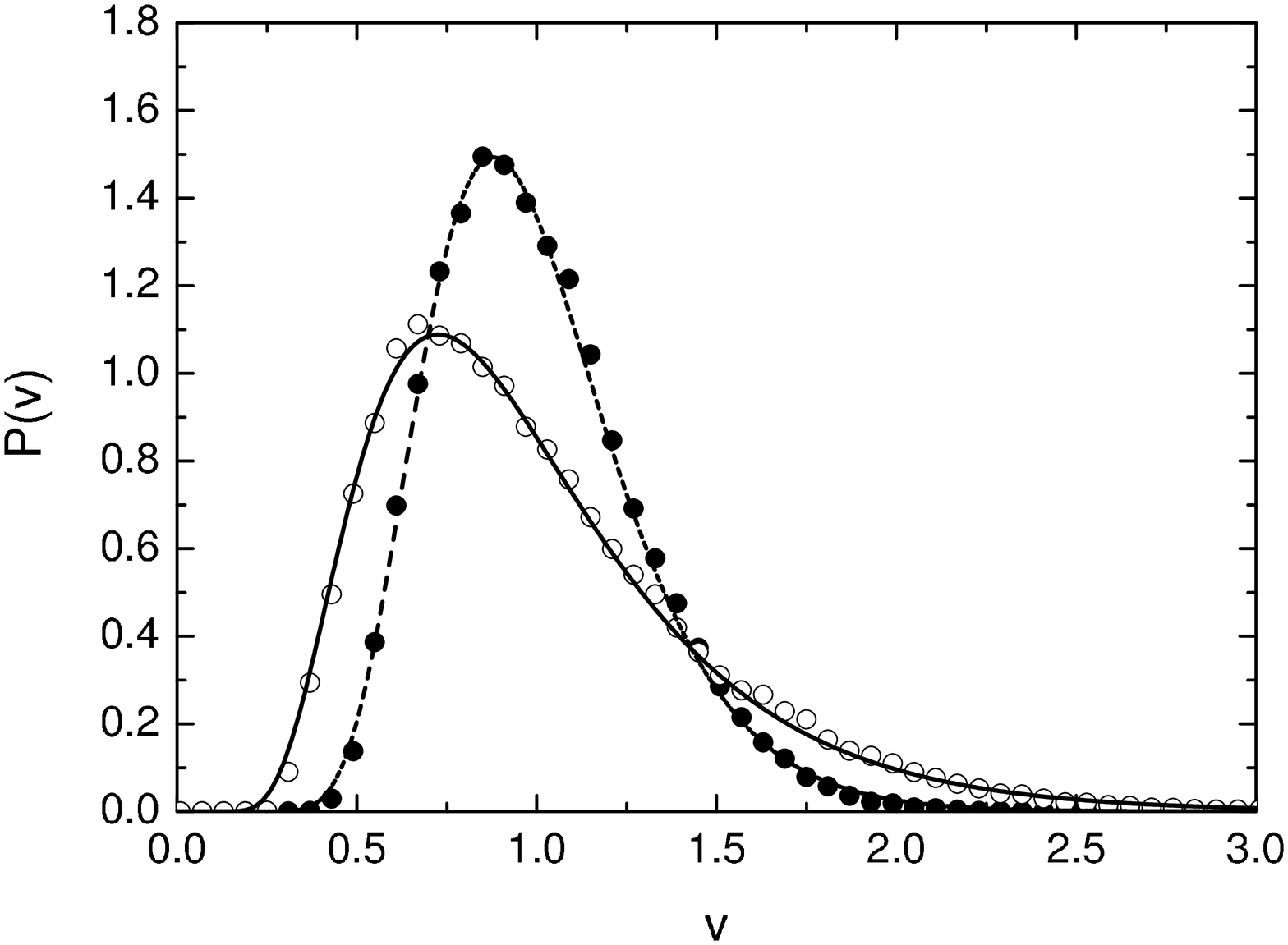}} \caption{Numerical
distribution $P(v)$ of the imaginary part of the $K$ matrix  for
the two values of the mean absorption parameter: $\bar{\gamma }=
19.3$ (open circles) and $\bar{\gamma } = 47.7$ (full circles),
respectively. The corresponding theoretical distribution $P(v)$
evaluated from the Eq.~(\ref{Equation34}) is marked by the solid
line ($\gamma = 19.3$) and dashed line ($\gamma = 47.7$),
respectively.}\label{Fig6}
\end{center}
\end{figure}

In Figure~5 the experimental distribution $P(v)$ of the imaginary
part of the $K$ matrix is shown for the two mean values of the
parameter $\bar{\gamma } = 19.3 \textrm{ and } 47.7$,
respectively. The distribution is the result of averaging over 88
and 74 realizations of the networks with the attenuators 1 dB and
2 dB, respectively. The experimental results in Figure~5 are in
general in good agreement with the theoretical ones. However, both
experimental distributions are slightly higher than the
theoretical ones  in the vicinity of their maxima.

\begin{figure}[!]
\begin{center}
\rotatebox{270} {\includegraphics[width=0.5\textwidth,
height=0.6\textheight, keepaspectratio]{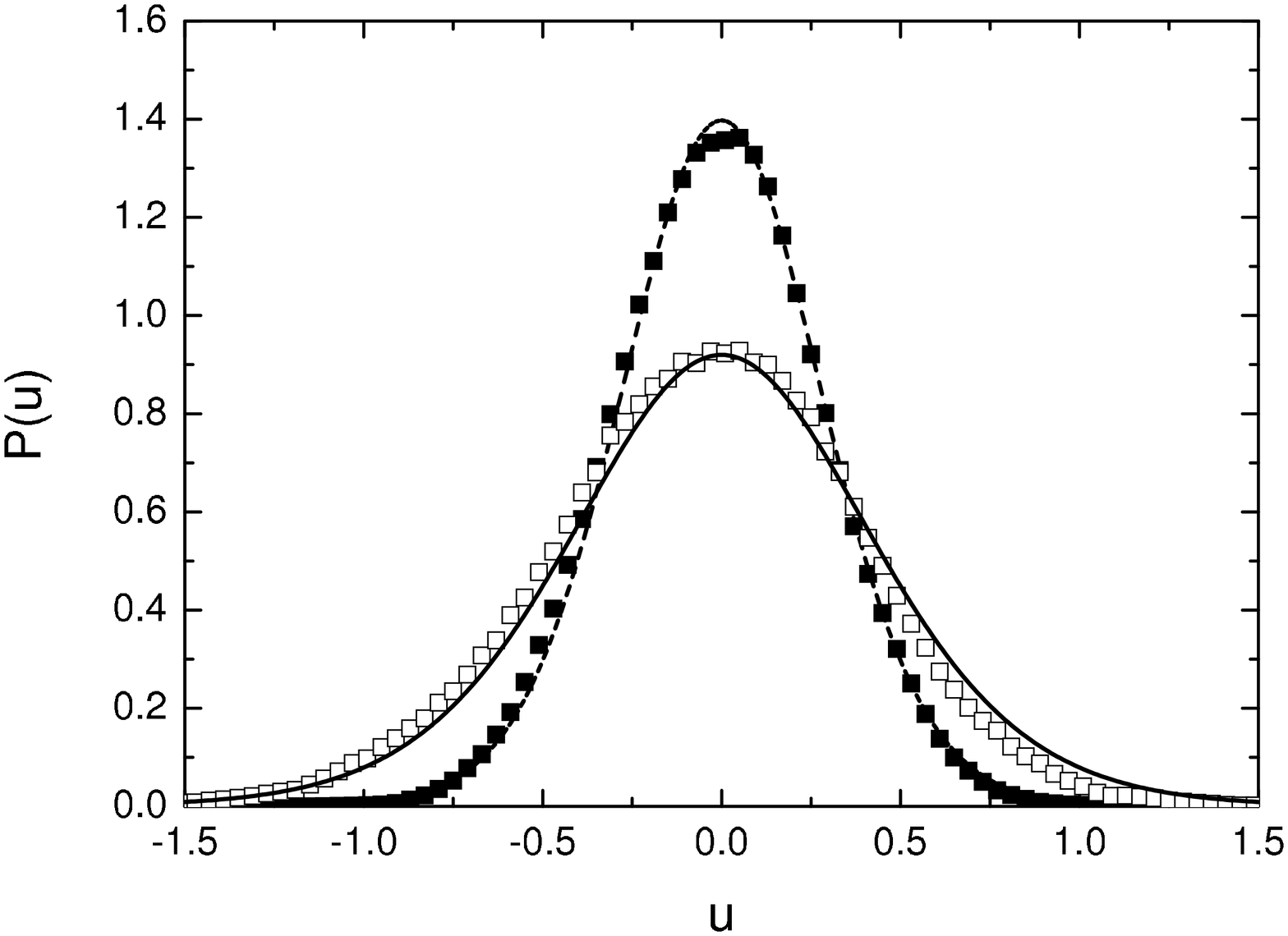}}
\caption{Experimental distribution $P(u)$ of the real part of the
$K$ matrix for the two values of the mean absorption parameter:
$\bar{\gamma } = 19.3$ (open squares) and $\bar{\gamma } = 47.7$
(full squares), respectively. The experiment is compared to the
theoretical distribution $P(u)$ evaluated from the
Eq.~(\ref{Equation35}): solid line ($\gamma = 19.3$) and dashed
line ($\gamma = 47.7$).}\label{Fig7}

\rotatebox{270} {\includegraphics[width=0.5\textwidth,
height=0.6\textheight, keepaspectratio]{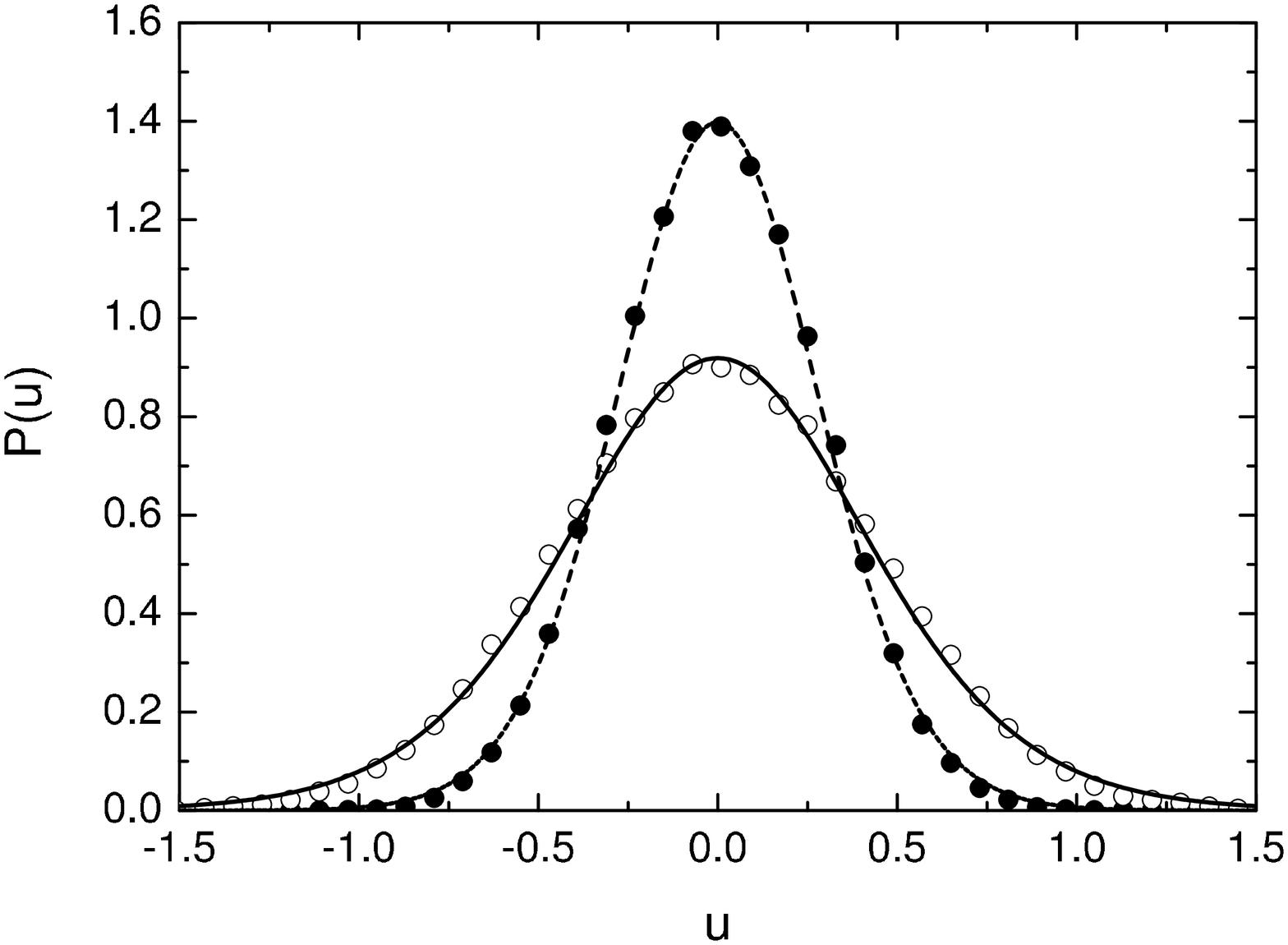}} \caption{Numerical
distribution $P(u)$ of the real part of the $K$ matrix for the two
values of the mean absorption parameter: $\bar{\gamma }=19.3$
(open circles) and $\bar{\gamma } = 47.7$ (full circles),
respectively. The numerical results are compared to the
theoretical distributions $P(u)$ evaluated from the
Eq.~(\ref{Equation35}): solid line ($\gamma = 19.3$) and dashed
line ($\gamma = 47.7$).}\label{Fig8}
\end{center}
\end{figure}

Results of the numerical calculations of the distributions $P(v)$
are shown in Figure~6 for  two mean values of the parameter
$\bar{\gamma }=19.3 \textrm{ and } 47.7$, respectively. They are
compared to the theoretical ones evaluated from the formula
Eq.~(\ref{Equation34}). Figure~6 shows also a good agreement
between the numerical and theoretical results, which confirms
usefulness of the optical potential approach in describing the
microwave attenuators.

Measurements of the distribution  $P(u)$ of the real part of the
Wigner's reaction matrix  give an additional test of the
consistency of the $\gamma$ evaluation. In Figure~7 we show this
distribution obtained for  two values of $\bar{\gamma }=19.3
\textrm{ and } 47.7$, respectively, compared to the theoretical
ones evaluated from the formula Eq.~(\ref{Equation35}). Also here
we observe good overall agreement between the experimental  and
theoretical results. However, Figure~7 shows that for the networks
with 2 dB attenuators the theoretical distribution is in the
middle ($-0.1<u<0.1$) slightly higher than the experimental one.
According to the definition of the $K$ matrix (see Eq. (1)) such a
behavior of the experimental distribution $P(u)$ suggests
deficiency of small values of $|\textrm{Im} S|$, whose origin is
not known. Moreover, the experimental distribution $P(u)$ obtained
for the networks assembled with 1 dB attenuators is slightly
asymmetric for $|u|>0.5$.

In Figure~8 the comparison of the numerical distribution $P(u)$
obtained for two values of $\bar{\gamma } = 19.3 \textrm{ and }
47.7$, respectively, to the theoretical one evaluated from the
formula Eq.~(\ref{Equation35}) is presented. In this case we see a
good overall agreement between the numerical and theoretical
results.

In spite of the above mentioned discrepancies which appeared mainly
in the case of the experimental distribution $P(u)$ the overall good
agreement between the experimental and theoretical results justifies
\textit{a posteriori} the chosen procedure of calculating the
experimental $\gamma$. The same is true also for the numerical
simulations.

In summary, we measured and calculated numerically the
distribution of the reflection coefficient $P(R)$ and the
distributions of the imaginary $P(v)$ and the real $P(u)$ parts of
the Wigner's reaction $K$ matrix for irregular fully connected
hexagon networks and graphs in the presence of strong absorption.
In the case of the microwave networks consisting of SMA cables and
attenuators the application of attenuators allowed for effective
change of absorption in the graphs. In the numerical calculations
absorption in an attenuator was modelled by an optical potential.
We showed that in the case of the time reversal symmetry
($\beta=1$) the experimental and numerical results for $P(R)$,
$P(v)$ and $P(u)$ are in good overall agreement with the
theoretical predictions. The agreement of the numerical and
theoretical results strongly confirms  the usefulness of the
optical potential approach in the description of the microwave
attenuators.

Acknowledgments. This work was partially supported by Polish
Ministry of Science and Higher Education grant No. N202 099
31/0746 and by the Ministry of Education, Youth and Sports of the
Czech Republic within the project LC06002.

\end{document}